\documentclass[apj]{emulateapj}
\newcommand{\pasa}{PASA}%
\newcommand{\nar}{New A Rev.}%
\newcommand{\Oiii}{[O\,\textsc{iii}]}

\newcommand{\Oi}{[O\,\textsc{i}]}
\newcommand{\Nii}{[N\,\textsc{ii}]}
\newcommand{\Sii}{[S\,\textsc{ii}]}

\newcommand{\Ariii}{[Ar\,\textsc{iii}]}
\newcommand{\Ha}{H$\alpha$}
\newcommand{\Hii}{H\,\textsc{ii}}
\newcommand{\Hei}{He\,\textsc{i}}
\newcommand{\Heii}{He\,\textsc{ii}}

\newcommand{\V}{\emph{V}}
\newcommand{\Rc}{\emph{R$_c$}}

\slugcomment{Received/Accepted}

\shorttitle{Trailing Streamers near LMC\,X-1}
\shortauthors{Cooke et al.}

\begin{document}

\title{Spectacular Trailing Streamers near LMC\,X-1:\\
    The First Evidence of a Jet?}

\author{Ryan Cooke\altaffilmark{1}, Zdenka Kuncic\altaffilmark{1},
Rob Sharp\altaffilmark{2} \& Joss Bland-Hawthorn\altaffilmark{2}}
\altaffiltext{1}{School of Physics, University of Sydney, NSW, Australia}
\altaffiltext{2}{Anglo-Australian Observatory, Epping, NSW, Australia}
\email{r.cooke@physics.usyd.edu.au}


\begin{abstract}
We report VIMOS integral field spectroscopy of the N159F nebula surrounding LMC\,X-1. Our observations reveal a rich, extended system of emission line filaments lining the boundary of a large conical cavity identified in \emph{Spitzer} mid-IR imaging. We find that X-ray photoionization cannot be solely responsible for the observed ionization structure of N159F. We propose that the extended filamentary emission is produced primarily by ionization from a shock driven by a presently unobserved jet from LMC\,X-1. We infer a shock velocity of $v_s\,\approx 90\,{\rm km\,s}^{-1}$ and conclude that the jet responsible for the bow shock is presently undetected because it has switched off, rather than because it has a low surface brightness. This interpretation is consistent with the present soft X-ray spectral state of LMC\,X-1 and suggests the jet is intermittent.


\end{abstract}

\keywords{accretion --- ISM: jets and outflows --- techniques: spectroscopic --- X-rays: binaries --- X-rays: individual (LMC\,X-1)}

\section{Introduction}

LMC\,X-1 resides in the highly complex nebula region N159
\citep*{hen56}. It is the most powerful known X-ray source in the LMC
\citep*{joh79}, with an X-ray luminosity $L_X \, \approx \,2 \times
10^{38} {\rm erg\,s}^{-1}$ \citep*{sch94}, and is believed to be
powered by accretion onto a stellar mass ($4-10 \,M_\odot$) black hole
with a companion O7\,\textsc{iii}-type star identified as star 32
\citep*{pak86,cow92}.
\citet{pak86} collated extensive spectral data and ruled out
collisional excitation by a supernova remnant. They concluded that
their detection of \Heii\ emission is the first evidence for an X-ray
ionized nebula (see also \citealt{ram06}).
We note, however, that the slit
spectra used in these studies  were limited to a relatively localized
region of the nebula, in the immediate vicinity of LMC\,X-1.

Using integral field spectroscopy (IFS), we present new evidence suggesting that X-ray photoionization cannot be solely responsible for the observed large scale morphology and high excitation lines, which we propose are best understood in terms of shock ionization attributable to an as yet undetected relativistic jet from LMC\,X-1. Indeed, the highly complex ionization structure of the extended nebula surrounding LMC\,X-1 is strikingly reminiscent of the other two only known examples of jet-induced shock ionized nebulae, those surrounding the powerful Galactic X-ray binaries SS433 \citep*{dub98} and Cygnus X-1 \citep*{gal05}.


We were initially alerted to the possibility of shock ionization in
N159F after observing with the Taurus Tunable Filter (TTF) at the
Anglo-Australian Telescope \citep{bhj98} in 2003 December.  We
discovered a rich system of streaming H$\alpha$ filaments in the
arcmin-scale nebula. The filaments trail back from the apex of a
conical cavity in the dust emission seen in the \emph{Spitzer}
Infrared Array Camera (IRAC) image of \citet{jon05}.
The apex is positioned $\approx 12''$ South-West of LMC\,X-1,  corresponding to $\approx 3.2\,{\rm pc}$ at a distance $d \approx 55 \, {\rm kpc}$ to the LMC \citep{Feast99}.

Motivated by these observations, we undertook further investigation into the environment of LMC\,X-1 through integral field spectroscopy (IFS) with VIMOS at the Very Large Telescope (VLT) in 2005 November. The new data reveal strong detections of optical emission lines due to \Oi, \Nii, \Sii, \Hei\ and \Ariii, as well as {\Ha}.
The emission lines all show a similar morphological bow shock structure as the initial \Ha\ filamentary streamers seen in the TTF image.
In Figure 3 of \citet{jon05}, which overlays the 5\,GHz radio contours of
\citet*{hun94} on an IR image of N159, radio emission appears to trace the IRAC cavity.





We summarize the observations and data reduction procedures in \S\,2. In \S\,3, we present our results and provide an interpretation of the environment of LMC\,X-1. Our conclusions are summarized in \S\,4.

\section{Observations and Data Analysis}

Our initial TTF observations of the N159F nebula were taken
in a $5$\,\AA\ bandpass centred 
on \Ha. A powerful differential technique called `straddle shuffling'
\citep*{mal01} was used to remove the surrounding continuum emission
to better than 1\% accuracy, revealing the filamentary structure of
the nebula. These data prompted observations with the VIMOS integral field unit (IFU) at the VLT on 2005 November 2, 5, 6.  Conditions were reported as photometric, with $1.4''-1.5''$ seeing (well matched to the $0.67''$ spatial scale of the IFU).  Observations were queue scheduled using the HR-Orange grating (R\,$\approx$\,2500, $\lambda\lambda$5250-7400\AA) and include the emission lines \Ha, \Oi, \Nii, \Sii, \Hei, and \Ariii. The systemic radial velocity ($v_r \approx 270 \, {\rm km\,s}^{-1}$) of the LMC ensures that the \Oi\ and \Ha\ emission lines are well separated from their telluric counterparts.

Given the $27''\times\,27''$ field-of-view of the VIMOS IFU in high resolution mode, four VIMOS pointings were required (with a 5\,pixel/$3.3''$ overlap) to tile the \Ha\ filaments along the boundaries of the Mid-IR cavity in the vicinity of LMC\,X-1. Three 1160\,sec exposures were taken at each of the four pointings. The data were reduced using the reduction pipeline VIPGI\footnote{VIPGI - Vimos Interactive Pipeline Graphical Interface, obtained from 
\emph{http://cosmos.iasf-milano.inaf.it/pandora/}} \cite*[see][for
  details]{sco05}. Datacubes were created and sub-sections of the
final mosaic were combined using a set of IFU data manipulation
routines created in IDL.
Emission line fitting was performed for each prominent spectral line using a single unresolved Gaussian profile.

Due to the restricted spectral coverage of the chosen VIMOS setting, the key shock diagnostic line \Oiii$\lambda$5007 is not present in our data. However, we retrieved archival \Oiii\ observations 
from the public ESO archive. The 2.3\,m WFI observations\footnote{Program ID: 076.D-0017(A)} of N159F consists of 2\,$\times$\,500\,sec observations using the \Oiii\ filter and 2\,$\times$\,125\,sec observations using the \Rc\ filter. The data were processed in the usual manner (overscan corrected, flatfielded from twilight flat frames, aligned and combined) using elements of the CASU imaging processing toolkit \citep*{irw01}. Continuum subtraction was achieved for the \Oiii\ and \Ha\ narrow band images using the aligned and scaled \V\ and \Rc\ images respectively. The WFI and VIMOS data were spatially registered and resampled using stellar images common to both data sets. 

\section{Results \& Discussion}

Figure~1 shows the main results from the new VIMOS, WFI and IRAC observations. Together, these provide
compelling evidence for shock ionization in the N159F nebula
surrounding LMC\,X-1. Fig.~1(a-d) show, respectively, the prominent
emission lines \Ha, \Ariii\ and \Oi\ in the VIMOS data cube, and
\Oiii\ from the continuum subtracted WFI data.  A bow shock
morphology, with trailing filaments, is evident in the
images. Fig.~1(e) shows a line diagnostics plot formed from a
pixel-pixel comparison of \Oiii/\Oi\ vs. \Ariii/\Sii\ where four
color-coded ionization regions show spatial coherence over the N159F
nebula (see Fig. 1(f)). Fig.~1(h) is a 3-color \emph{Spitzer} IRAC
image centered on star 32 \citep{jon05}, showing a dust cavity
associated with the proposed bow shock. Fig.~1(k) presents a 3-color
composite image in \Nii\ (red), \Ha\ (green) and \Oi\ (blue).
We speculate that the shock is driven by a jet, with orientation indicated in Fig.~1(k).

\subsection{Disentangling the Shock-Ionized Nebula}


Separating out the shock-ionized part of the optical nebula is not straightforward, as radiative shocks can produce collisionally excited optical emission lines similar to those produced by X-ray photoionization \citep*[see e.g.][]{dop03}. Both ionization mechanisms can produce enhanced (relative to \Hii\ regions) \Ha\ as well as enhanced forbidden lines of neutral or low ionization species, such as \Oi$\lambda$6300, \Sii$\lambda\lambda$6716,6731 and \Nii$\lambda\lambda$6548,6583, and of higher ionization species, such as \Oiii$\lambda\lambda$5007,4959 and \Ariii$\lambda$7136. \Heii\,$\lambda$4686 is a reliable X-ray photoionization diagnostic as it is difficult to produce with shock ionization, while \Ariii\ is difficult to produce with X-ray photoionization, except for very high ionization parameters (R. Sutherland, private communication). Furthermore, X-ray photoionization tends to favour the production of \Oiii\ over \Oi\ \citep*{dra93}. On the other hand, \Oiii\ can be a powerful shock diagnostic, since all radiative shock models \citep*{cox85,bin85,har87} consistently predict a dramatic onset of \Oiii\ emission near a critical shock speed, $v_{\rm s}\simeq$\,100\,{\rm km\,s}$^{-1}$.

Whilst the \Ha\ and \Ariii\ images in Fig.~1(a) and (b) exhibit a
similar morphological structure suggestive of a bow shock, a
comparison of the \Oi\ and \Oiii\ images in Fig.~1(c) and (d) shows
regions of anticorrelation in the immediate vicinity of LMC\,X-1 and
around the B5\,\textsc{i} star R148 \citep*{fea60}, where there is a deficit of \Oi\ emission, but enhanced \Oiii\ emission. R148 is not capable of producing this high excitation emission. Indeed, its Str\"omgren sphere ($r_{s}\approx0.04\, {\rm pc}$) does not even encompass one pixel of our image.
The Str\"omgren radius for star 32 (the O7\,\textsc{iii} companion to
LMC\,X-1 -- \citealt{cui02}) is $\approx 0.9 \, {\rm pc}$, comparable to the size of the blue region defined in Fig.~1(e) and (f). Recalling that X-ray photoionization favours \Oiii\ over \Oi\ and can produce \Ariii\ for high ionization parameters, this blue region, which exhibits the largest local enhancement in \Oiii/\Oi\ and \Ariii/\Sii, must be due to the combined effects of X-ray ionization by LMC\,X-1 and photoionization by star 32.
The yellow region in Fig.~1(e) and (f) is consistent with the
extent of the X-ray Str\"omgren radius ($\simeq 2\,{\rm pc}$), calculated by \citet{pak86}. Note, however, that this yellow region is located to the East (left) of R148, on the far side of LMC\,X-1 and spatially coincides with some of the streamers. Thus, we interpret the yellow zone of enhanced \Ariii/\Sii\ and \Oiii/\Oi\ as being due to a combination of shock ionization and X-ray ionization.
The red zone in Fig.~1(e) and (f) is predominantly shock-ionized. Fig.~1(i) and (j) show a proposed trail of streamers through this shock-ionized region.

The evidence for X-ray photoionization in N159F has hitherto been based on slit spectra of \Heii\,$\lambda$4686 emission detected in the immediate vicinity of LMC\,X-1 \citep{pak86,ram06}.
In Fig.~1(l), we show the positioning of the echelle slits used by \citet{ram06} to detect the \Heii\ emission. \citet{pak86} used similarly placed orthogonal slits, but centred on star 32 instead of R148. Note that although the echelle slits coincide with the yellow zone we identify as being both shock ionized and X-ray ionized, the observed \Heii\,$\lambda$4686 emission is almost certainly due to only X-ray ionization as none of the radiative shock models predict significant \Heii\,$\lambda$4686. Note also that it would have been extremely difficult to find unequivocal evidence for shock ionization from other line diagnostics due to the confusion with X-ray ionization in this region and also to the spatially limited extent of the slits.
On scales of several parsecs, our IFU data reveal enhanced \Oi/\Ha, \Oiii, and \Ariii, as well as \Ha, forming a shell-like structure around LMC\,X-1, but extending beyond the X-ray Str\"omgren radius (that is, beyond the yellow zone in Fig.~1(e) and (f)). This is compelling evidence that shock-ionization rather than X-ray ionization is primarily responsible for this high-excitation emission seen on the largest scales in the nebula.



\subsection{Physical parameters}
Fig.~1(g) shows an image of the electron number density, $N_e$, deduced from the \Sii\ emission \citep*{ost06}. It reveals an enhanced flattened region that falls slightly inside and near the apex of the bow-shock morphology. This structure is highly suggestive of a `Mach disk' commonly seen in bow shocks associated with Herbig-Haro jets \citep{har99}. In the flattened region, we find $N_e \! \approx \! 1600 \, {\rm cm}^{-3}$, a factor $\approx \! 3$ times greater than the environment.
This is less than the maximum compression ratio of $4$ expected for a strong shock, but is compatible with the Mach number inferred from the shock speed (see below).

From the high density region, we can determine a shock speed using the
multitude of line ratios extracted from a single IFS exposure. Fig.~1(m) shows theoretical line ratios from the radiative shock model of \citet{har87}. The dashed horizontal lines indicate the maximum and minimum shock speeds obtained from the new data for each corresponding line ratio.  From the presence of \Oiii, we infer a lower limit $v_s$ $\gtrsim$ 80 km\,s$^{-1}$. \Ariii/\Hei\ suggests $v_s$ $\lesssim$ 190\,km\,s$^{-1}$, while 
both the \Sii/\Ha\  and \Ariii/\Ha\  line ratios further constrain the shock velocity to $v_s \! \approx 90-100 \,{\rm km \, s}^{-1}$.
Thus, we estimate a shock speed $v_s\approx$\,90\,km\,s$^{-1}$, as indicated by the black dashed vertical line in Fig.~1(m). This is broadly consistent with the line profiles being marginally resolved at $\approx 120-150$ km s$^{-1}$ FWHM within the high density region. It also implies a Mach number of $\approx \! 6$, assuming an isothermal sound speed $\approx \! 14\,{\rm km \, s}^{-1}$ for an ambient gas temperature of $10^4 \, {\rm K}$. We note, however, that the deduced value of $v_s$ is likely to be somewhat overestimated, as the \citet{har87} shock model assumes a neutral pre-shock medium, and this produces a spectrum similar to preionized, shocked material with a lower $v_s$ \citep{cox85}.

\subsection{Energy Budget}
The streamers are unlike the uniform wind-blown bow shocks associated
with OB-runaway stars \citep*[see e.g.][]{kap97}.
Indeed, the the inferred space velocity $v_\star$ of the
O7\,\textsc{iii} companion is too low to be consistent with this
interpretation.
We estimate $v_{\star}\,\sim\,$0.4$\, (\dot M_w / 10^{-7}\, M_\odot \,
{\rm yr}^{-1})^{1/2} \, v^{1/2}_{w,1500} \, (\rho_a /
6\times\,10^{-22}\,{\rm g \, cm^{-3}})^{-1/2} \, R^{-1}_{3.2} \, {\rm
  km \, s}^{-1}$ \citep{kap97}, where $v_w \, = \, 1500 \, v_{w,1500}$
and $\dot M_w$ are the wind velocity and mass loss rates of star 32
\citep{pak86}, $\rho_a$ is the mass density of the ambient medium,
deduced from the \Sii\ lines (\S\,3.2), using an ionization fraction
of $x \, \approx\, 1$, and where $R\,=\,3.2\,R_{3.2}$ is the distance
between star 32 and the stagnation point (apex), which has an angular
distance $12''$ corresponding to $\gtrsim\,3.2\,{\rm pc}$, where the
inequality takes into account projection effects.
We henceforth consider the possibility that the shock is produced by a jet from LMC\,X-1.

Adopting the ``dentist drill'' model of \citet{Scheuer74}, a cavity,
or cocoon, develops naturally around a jet ploughing through the
ambient ISM.
A shock develops around the cocoon and the shocked gas then comes into dynamical (ram) pressure equilibrium with the jet. Thus,
$ \rho_{j}v_{j}^2 \approx \rho_{s}v_{s}^2 \, ,$
where $\rho_j$ and $\rho_s$ are the mass densities of the jet and shocked gas, respectively, and $v_j$ and $v_s$ are the corresponding speeds. The total jet kinetic power is $P_j \! \approx \! \pi (\phi_j z_j)^2 \rho_j v_j^3$ (assuming it is not too relativistic), where $\phi_j$ is the jet half opening angle and $z_j$ is the jet length (distance to the impact site, at the apex of the cocoon, from LMC\,X-1). Thus,
$
P_j \simeq 2 \times 10^{39} \, v_{j,0.1} \,  \left( \phi_{j,10} \, z_{j,3.2} \, v_{s,90} \right)^2
( \rho_s / 2 \times 10^{-21} \, {\rm g \, cm^{-3}} ) \, {\rm erg \, s^{-1}} \, .
$
Here, $\phi_{j,10} = \phi_j/10^\circ$ (\citealt*{M-JFN06} find  an
upper limit of $\phi_j \lesssim 10^\circ$ for X-ray binary jets),
$z_{j,3.2} = z_j\sin^{-1}\theta/(3.2\,{\rm pc})$ and $\theta$ is the
jet inclination angle to our line-of-sight.
We have used a post-shock electron number density $N_e \approx 1.6
\times 10^3 \, {\rm cm}^{-3}$
(\S\,3.2) to estimate $\rho_s$.
This estimate of jet power falls squarely within the range $P_j
\approx 10^{37 -42} \, {\rm erg \, s^{-1}}$ estimated by
\citet{FendPool00} for quasi-steady X-ray binary jets.

It is noteworthy, however, that LMC\,X-1 currently appears to be in a
persistent high/soft X-ray spectral state \citep{sch94}, during which
it is unlikely to be producing a powerful jet \citep*{fen04} and its
X-ray luminosity is dominated by the accretion disk, which radiates away all the accretion power. Since the accretion power inferred from the observed X-ray luminosity, $\approx 2 \times 10^{38} \, {\rm erg \, s^{-1}}$, is an order of magnitude less than the inferred jet power, we suggest that rather than being a dark jet (as \citealt{gal05} proposed for  Cyg\,X-1, which is in a low/hard spectral state), the non-detection of the LMC\,X-1 jet results from the inherent transient nature of X-ray binary jets. The jet is not presently seen because it may have recently switched off and the shock-ionized nebula surrounding LMC\,X-1 is still radiating the energy impacted during an earlier jet-active epoch. We estimate a radiative cooling timescale of just $\sim 10^{2}\,$yrs, assuming $0.3$ solar metallicity \citep{dop03}. The jet-active phase may have been triggered by a relatively sudden increase in the mass accretion rate, which has a present value $\dot M_a \approx 12 L_X/c^2 \approx 4 \times 10^{-8} \, M_\odot \, {\rm yr}^{-1}$ (assuming a radiative efficiency of $\frac{1}{12}$ for standard thin disk accretion). The dynamical timescale over which the jet-driven shock has been expanding into the ISM is estimated as $t_{dyn} \sim z_j / v_s \gtrsim 0.03\,$Myr, which is remarkably similar to jet lifetime estimates obtained for other X-ray binaries \citep{gal05,Kaiser04}.

\section{Conclusions}

We have demonstrated the extraordinary power of integral field
spectroscopy in our study of the emission-line nebula N159F around
LMC\,X-1. Different emission-line diagnostics have been combined to
separate out different ionization regions on extended scales.  We
propose that three ionizing sources contribute to the complex nature
of N159F: 1. photoionization by star 32 (the O7\,\textsc{iii}
companion to LMC\,X-1); 2. X-ray photoionization by LMC\,X-1 itself;
and 3. jet-driven shock ionization. The first two phenomena are
relatively localized, whereas shock ionization dominates over the
extended streamers, which appear to be bounded by a bow-shock
morphology
The morphology and ionization properties of the extended nebula around LMC\,X-1 are strikingly similar to those seen in other known jet-driven shock ionized nebulae around Galactic X-ray binaries.
We infer a shock speed of $v_s \!\approx \! 90$ km s$^{-1}$
and deduce the power of the jet driving the shock to be $\approx\,2\times 10^{39}$ erg s$^{-1}$. Since LMC\,X-1 is unlikely to be producing a jet in its current X-ray spectral state, we propose that the jet is intermittent, with a characteristic lifetime of $\sim 0.03\,$Myr.

\acknowledgments
We thank an anonymous referee whose comments and suggestions improved the paper considerably. We also thank R. Sutherland, M. Dopita, B. Gaenslar, K. Blundell, and R. Soria for valuable discussions.

\begin{figure}
\centerline{\includegraphics[width=16.5cm]{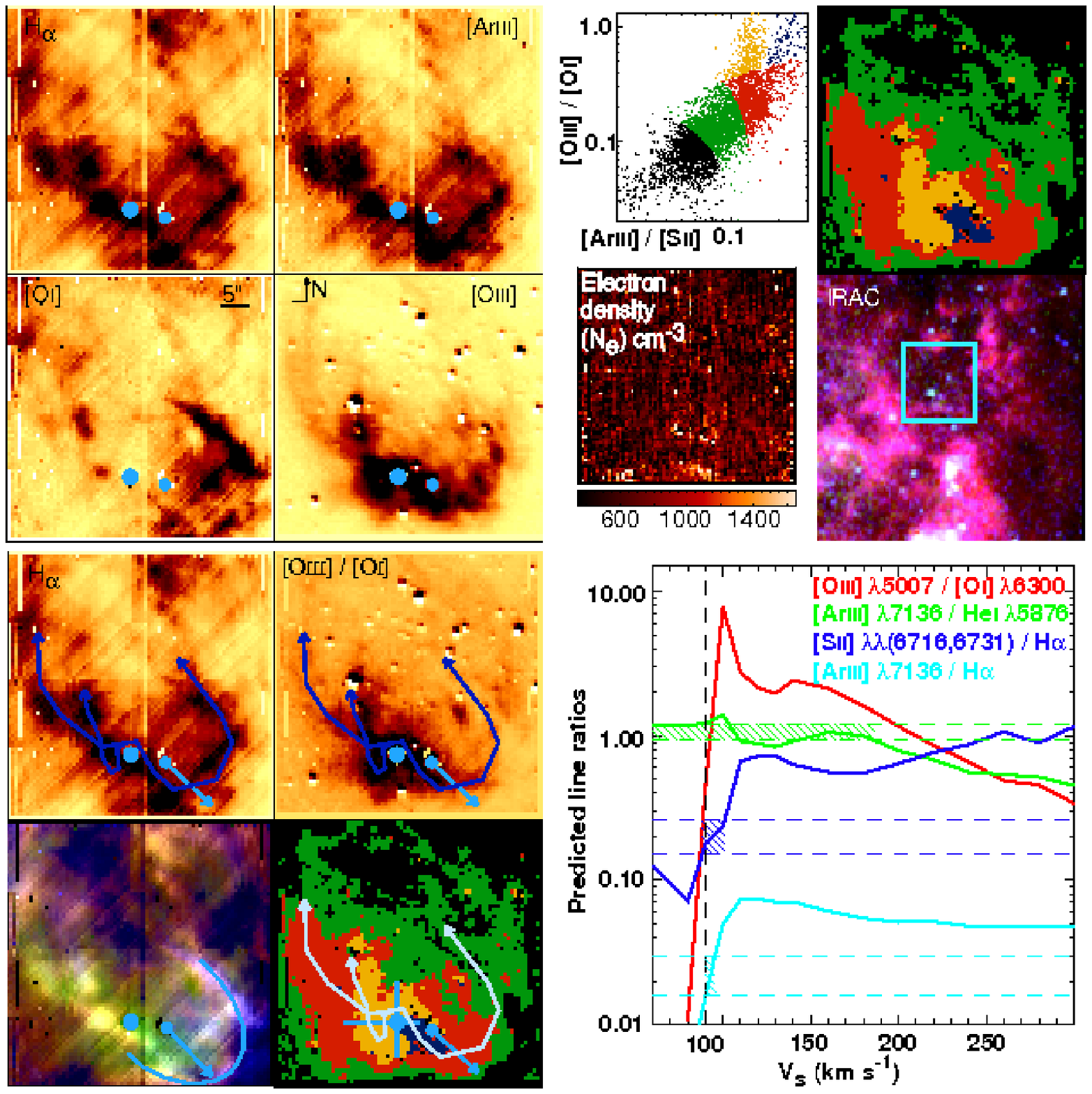}}
\caption{(a)--(d) Continuum subtracted integrated line flux images of N159F in \Ha, \Ariii, \Oi, and \Oiii; filled blue circles indicate the locations of stars 32 (right) and R148 (left); north is up and east is to the left ($5''\approx1.3\, {\rm pc}$);
(e) A diagnostic scatterplot of \Oiii$\lambda$5007/\Oi$\lambda$6300 versus \Ariii $\lambda$7136/\Sii $\lambda$(6716+6731); the colour coding corresponds to different ionization regions shown in (f);
(g) An electron density image, showing an arc of higher electron density to the south;
(h) The \textit{Spitzer} three colour composite image of N159F
  \citep{jon05},
constructed from the IRAC 3.6, 4.5 and 8.0\,$\mu$m bands;
(i)--(j) \Ha\ and \Oiii/\Oi\ images overlaid with a proposed trail of the streamer filaments;
(k) A false-colour image (R,\,G,\,B = \Nii$\lambda$6583,\,\Ha,\,\Oi$\lambda$6300) with the proposed bow shock and jet orientation indicated;
(l) Same as (f), with overlaid blue lines indicating the echelle slit
positions of \citet{ram06}; grey lines are the proposed streamers; 
(m) Predicted line ratios and corresponding shock speeds from the radiative shock model in \citet{har87} (solid curves) overplotted with observed ranges of line ratios (dashed lines) measured from the region of high electron density shown in (g).
\label{fig1}
}
\end{figure}

\end{document}